# Bioinspired Synergistic Texture and Color Modulation Enabled by Surface Instability of Cholesteric Liquid Crystal Elastomers

Xiao Yang[1]†, Jay Sim[1]†, Wenbin Huang[1], Ruike Renee Zhao[1]*

**Affiliations**
[1]Department of Mechanical Engineering, Stanford University, Stanford, CA 94305, USA.
*Corresponding author. Email: rrzhao@stanford.edu (R.R.Z.)
†These authors contributed equally to this work.

**Abstract:** Certain cephalopods can dynamically camouflage by altering both skin texture and color to match their surroundings. Inspired by this capability, we present a cholesteric liquid crystal elastomer-liquid crystal elastomer (CLCE-LCE) bilayer capable of simultaneous, reversible modulation of surface texture and structural color through programmable wrinkling. By tuning the bilayer's fabrication parameters, on-demand wrinkle morphologies and color combinations are achieved. Spatially selective UV curing allows localized surface textures, while chemical patterning of the CLCE layer enables region-specific color responses, expanding the design space for multifunctional, spatially encoded optical materials. The CLCE-LCE bilayer enables dynamic thermal regulation by tuning light absorption through synergistically modulating surface morphology and color. Notably, this system achieves strain-dependent multistate encoding via multistep selective UV curing, revealing distinct visual content under different applied strains. This work establishes a versatile platform that merges surface instabilities with tunable structural coloration, advancing intelligent materials with programmable, strain-responsive surface and optical properties.

## INTRODUCTION

Through natural evolution, various species developed sophisticated bodily functions that allow them to dynamically change their skin color for camouflage, signaling, and reproduction (*1, 2*). Inspired by these animals, responsive materials capable of altering their colors in response to external stimuli (*3, 4*) have been developed. Among these systems, mechanochromism (*5, 6*) can be readily achieved through mechanical deformation of the periodic nanostructures of structurally colored materials, such as photonic crystals (*7, 8*) and liquid crystals (*9, 10*). These mechanochromic materials have found applications in strain sensors (*11, 12*), wearable electronic skins (*13, 14*), and optical encryption (*15*). However, most existing mechanochromic materials only demonstrate color-changing functionality.

Certain organisms, like octopuses and cuttlefish of the cephalopod class, exhibit improved camouflage abilities by changing their skin texture in addition to skin color (**Fig. 1A**). These remarkable abilities rely on their advanced skin structures capable of manipulating light transmission, absorption, and reflection (*16*). An artificial material system integrating such nature-inspired texture and color change would expand material functionalities and advance various applications such as adaptive camouflage and display devices (*17-19*). One strategy to induce texture change is wrinkling, a surface instability-driven phenomenon that is ubiquitous in natural systems, such as aging human skin and dried fruit (*20-23*). Prior research has focused on wrinkling systems capable of dynamically tuning texture and optical properties by forming or releasing surface wrinkles (*24-27*). Only a limited number of studies have investigated coupled texture and color change in wrinkling systems. For example, a nanoscale wrinkled surface can exhibit structural color when the wrinkle periodicity falls within the visible light wavelength range (~400-700 nm) due to the optical interference effect (*28*). Applying strain to the wrinkled surface adjusts the wrinkle periodicity, leading to a shift in the reflection wavelength and thus, structural color change (*29*). However, the fabrication of such wrinkling systems relies on templates with predefined nanoscale wrinkle patterns, inherently limiting the tunability and programmability of texture and color change. Wrinkling can be engineered by inducing mismatch strains in a bilayer system composed of a stiff thin film bonded to a soft elastomer substrate (*30*). Using this approach, dynamic color change has been demonstrated in bilayer elastomers integrated with mechanochromic photonic crystals, where stress-induced micro-wrinkling modulates the lattice spacing of the photonic crystals (*31, 32*). However, the presence of rigid photonic crystals often leads to cracking and delamination. Moreover, as wrinkles form, increased light scattering from the micro-wrinkled surface diminishes color visibility due to the reduced reflectivity and the brightness of the structural color (*33*). These limitations highlight the urgent need for a system capable of programmable and effective tuning of surface texture and color, while maintaining robust mechanical reversibility, reliability, and high structural color visibility for advanced optical applications.

In this work, we design and fabricate a bioinspired cholesteric liquid crystal elastomer-liquid crystal elastomer (CLCE-LCE) bilayer capable of simultaneous and reversible changes in surface texture and structural color via wrinkling. By cooperatively controlling the CLCE film thickness, substrate pre-stretch ratio, bilayer modulus ratio, and initial CLCE color, bilayers with customizable wrinkle morphologies and color combinations are achieved. Finite element analysis (FEA) simulations are conducted to predict and guide the design of the wrinkle patterns and corresponding color distributions. The CLCE-LCE bilayer can achieve spatially programmable surface textures and color responses by localized UV curing and spatial chemical patterning of the CLCE film, enabling the integration of multiple surface textures and color modes within a single bilayer. The synergistic texture and color modulation is further demonstrated for applications in dynamic thermal regulation and multistate information encoding, highlighting the versatility of the CLCE-LCE bilayer. 2D texture and color modulation are also explored. It is anticipated that this work will establish a versatile platform for developing advanced multifunctional materials with programmable texture and color as well as deepen the understanding of the interplay between mechanical instabilities and structural color modulation in soft material systems.

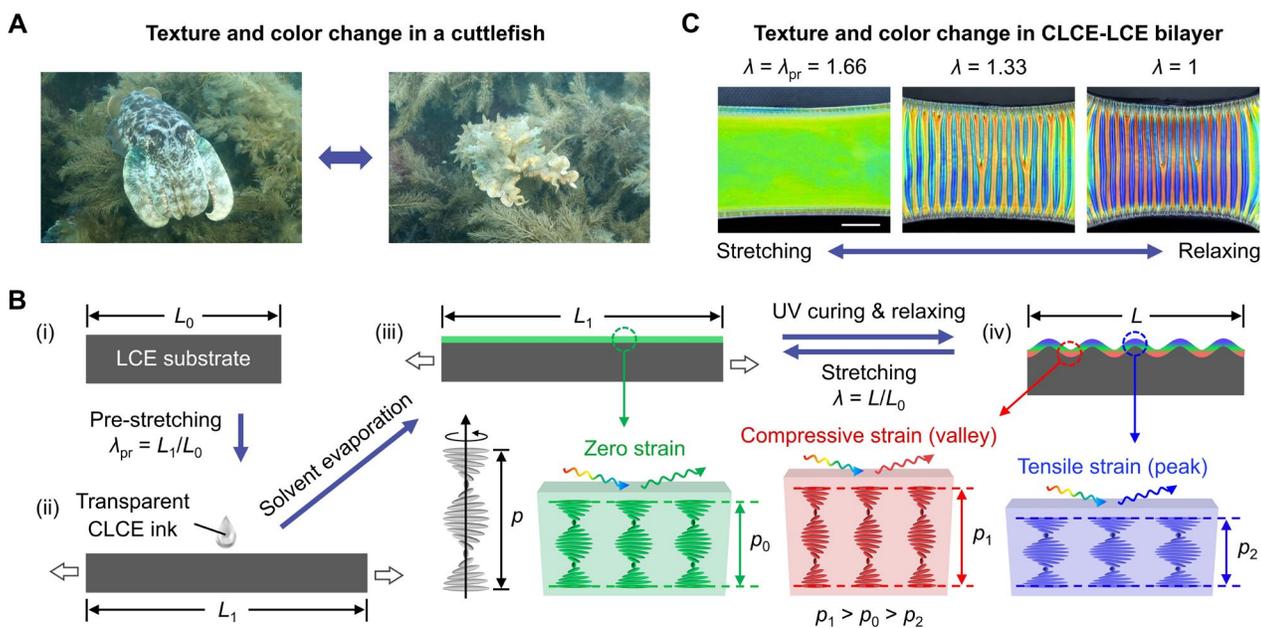

**Fig. 1. CLCE-LCE bilayer enabling synergistic texture and color changes via controllable wrinkling.** (**A**) Bioinspiration from a cuttlefish, which simultaneously changes its skin texture and color (image credit to Roger T. Hanlon, Marine Biological Laboratory, USA). (**B**) Schematics of CLCE-LCE bilayer mechanism showing coupled texture and color changes during wrinkling,

induced by mismatch strain and modulus ratio between the LCE substrate and CLCE film. The periodic tensile and compressive strain distribution across the wrinkled surface modulates the local helical pitch of the CLCE, leading to a spatially varying structural color. (**C**) Experimental images showing the evolution of texture and color in CLCE-LCE bilayer during the strain relaxation process. Scale bar: 5 mm.

## RESULTS

### Fabrication and characterization of the CLCE-LCE bilayer

**Figure 1B** illustrates the fabrication process of the CLCE-LCE bilayer and the mechanism driving the coupled changes in surface texture and color via wrinkling. Wrinkling is engineered by inducing mismatch strains in a bilayer system. Here, a lightly crosslinked black dye-doped LCE substrate is first fabricated with a length of $L_0$ (**Fig. 1B**,i), and then pre-stretched to $L_1$ (**Fig. 1B**,ii). The ratio of the pre-stretched length $L_1$ to the original length $L_0$ is defined as the pre-stretch ratio $\lambda_{pr}$. The transparent CLCE ink is dropped onto the surface of the pre-stretched LCE substrate. As the solvent in the CLCE ink evaporates, the liquid crystal molecules self-assemble into periodic helical nanostructures that selectively reflect specific wavelengths of light, forming a stress-free, colored CLCE film as seen in **Fig. 1B**,iii. The structural color is governed by the helical pitch $p$, which is defined as the distance along the helix axis where the liquid crystal director completes a full 360° rotation (*34*). The CLCE layer is then cured onto the pre-stretched LCE substrate under UV irradiation, and forms a densely crosslinked network with significantly higher stiffness than the substrate. The moduli of both layers can be tuned by modifying their chemical compositions. The specific formulations of the CLCE and LCE are provided in **fig. S1**, and additional details on the fabrication process are available in **Materials and Methods**. Upon relaxing the pre-stretch-induced strain in the substrate, simultaneous texture and color changes appear on the CLCE film surface due to the wrinkling and the deformation-induced modulation of the initial pitch of $p_0$ (**Fig. 1B**,iv). The mismatch strain between the pre-stretched LCE substrate and stress-free CLCE film triggers surface instability, leading to the formation of a well-controlled wrinkle pattern. Concurrently, the periodic tensile and compressive strain distribution across the wrinkled surface modulates the local helical pitch of the CLCE, leading to a spatially varying structural color. At the wrinkle peaks, tensile strain reduces the pitch from $p_0$ to $p_2$, causing a blue shift. In contrast, the valleys experience compressive strain, increasing the pitch from $p_0$ to $p_1$, and resulting in a red shift (*35*). In addition, the robust interfacial bonding ensures the structural integrity of the bilayer during the strain relaxation process (see **figs. S2** and **S3** in Supplementary Materials for the tough interfacial bonding characterization).

**Figure 1C** shows the experimental relaxing process of a CLCE-LCE bilayer, with Young's moduli of 1.24 MPa and 0.05 MPa, respectively (see **fig. S4** in Supplementary Materials for material characterization). Initially, the LCE substrate (30 mm × 20 mm × 4 mm) is stretched along its length to a pre-stretch ratio $\lambda_{pr} = 1.66$. A green CLCE film is then formed on the pre-stretched LCE substrate. As the bilayer is relaxed from a stretched state ($\lambda = 1.66$, $\lambda$ is defined as the ratio of the current length $L$ to the original length $L_0$) back to its relaxed state ($\lambda = 1$), the bilayer undergoes a coupled wrinkling and color transformation. The wrinkling peaks (under tension) and valleys (under compression) exhibit blue and red colors, respectively (see **movie S1** in Supplementary Materials for the stretching-relaxing process of CLCE-LCE bilayer). This strain-induced texture and color change is fully reversible even after significant stretching-relaxing cycles, demonstrating excellent reversibility and reliability of the CLCE-LCE bilayer.

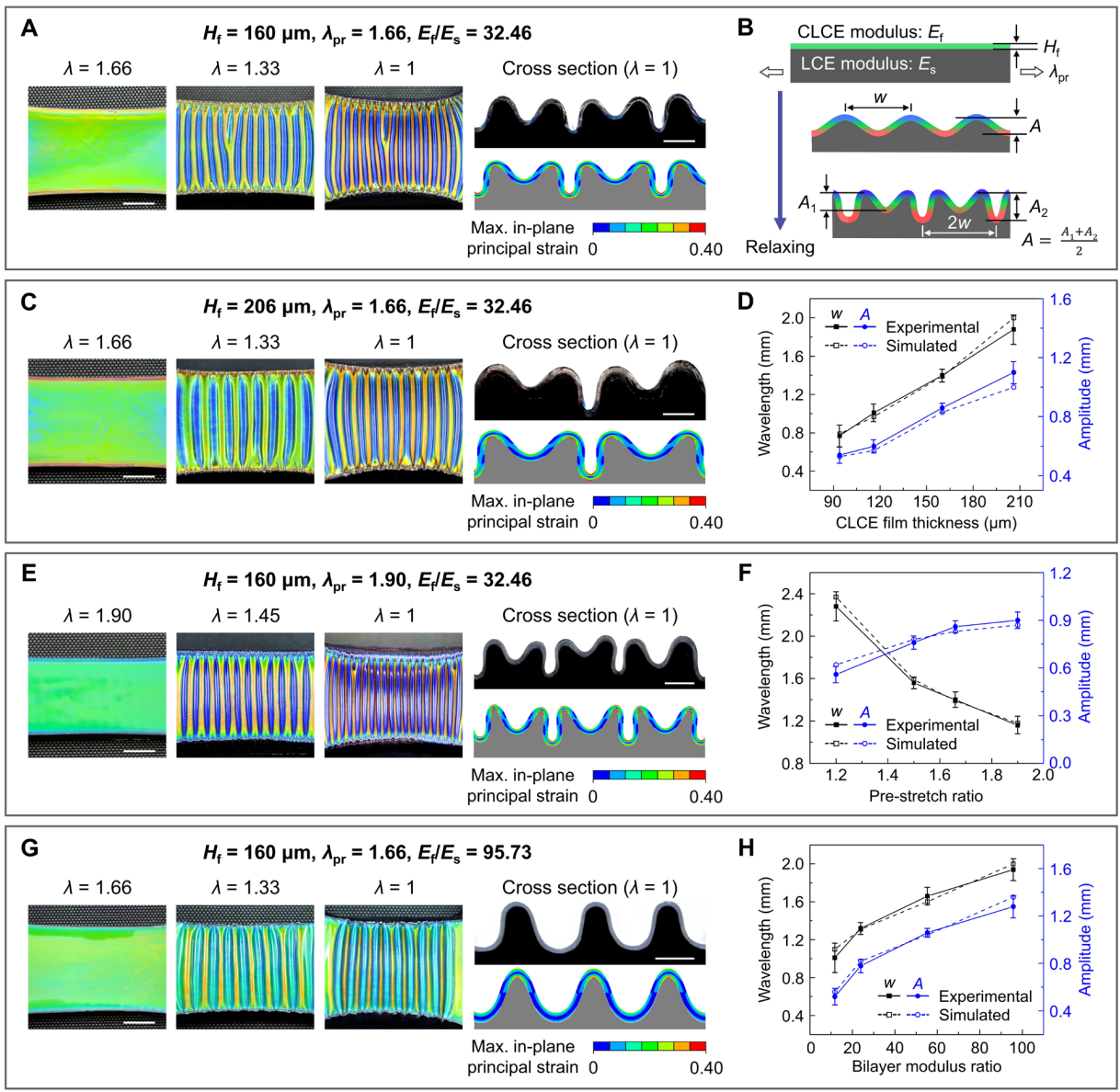

**Fig. 2. Effect of CLCE film thickness, pre-stretch ratio, and bilayer modulus ratio on texture and color modulation.** Sequential experimental images showing the evolution of texture and color in CLCE-LCE bilayers with **(A)** $H_f = 160$ μm, $\lambda_{pr} = 1.66$, $E_f/E_s = 32.46$, **(C)** $H_f = 206$ μm, $\lambda_{pr} = 1.66$, $E_f/E_s = 32.46$, **(E)** $H_f = 160$ μm, $\lambda_{pr} = 1.90$, $E_f/E_s = 32.46$, and **(G)** $H_f = 160$ μm, $\lambda_{pr} = 1.66$, $E_f/E_s = 95.73$ during the strain relaxation process, and corresponding experimental and FEA cross-sectional images at the fully relaxed state ($\lambda = 1$). **(B)** Schematics of characteristic parameters of the bilayer wrinkling system. Wavelength (black) and amplitude (blue) of fully relaxed CLCE-LCE bilayers as functions of **(D)** $H_f$, **(F)** $\lambda_{pr}$, and **(H)** $E_f/E_s$. Data are shown as mean ± s.d. ($n = 5$). Scale bars: 5 mm for top views and 1 mm for cross sections.

## Surface texture and color modulation by programming CLCE film thickness, LCE substrate pre-stretch ratio, and bilayer modulus ratio

In a bilayer wrinkling system, three key parameters, film thickness $H_f$, substrate pre-stretch ratio $\lambda_{pr}$, and bilayer modulus ratio $E_f/E_s$, govern the resulting wrinkle morphology (*36, 37*). This section investigates how these parameters influence the texture and color modulation of CLCE-LCE bilayers. **Figure 2** (**A and C**), (**A and E**), and (**A and G**) compare the effects of varying $H_f$, $\lambda_{pr}$, and $E_f/E_s$, respectively, on wrinkle evolution. **Figure 2B** defines the bilayers' initial parameters and the peak-to-peak wavelength $w$ and peak-to-valley amplitude $A$ measured at different wrinkling stages during the relaxation process. The baseline case (**Fig. 2A**) has $H_f = 160$ μm, $\lambda_{pr} = 1.66$, and $E_f/E_s = 32.46$. The variations include: **Fig. 2C**: $H_f = 206$ μm, with $\lambda_{pr}$ and $E_f/E_s$ unchanged; **Fig. 2E**: $\lambda_{pr} = 1.90$, with $H_f$ and $E_f/E_s$ unchanged; **Fig. 2G**: $E_f/E_s = 95.73$, with $H_f$ and $\lambda_{pr}$ unchanged. Experimental and FEA-predicted wrinkle cross-section profiles are shown for the fully relaxed state ($\lambda = 1$). The FEA visualizations are colored by the maximum in-plane principal strain distribution, illustrating strong agreement between simulation and observation. Note that the onset wrinkling wavelength follows the scaling law described by Cao et al. (*38*). See the section of "Wrinkle wavelength prediction and FEA implementation" in Supplementary Materials for more details. The corresponding color distribution across the CLCE film surface is also predicted by FEA, where the color mapping is based on the color-strain relationship from a CLCE layer under homogeneous strain (see **fig. S5** in Supplementary Materials for the color-strain relationship). The simulated color patterns show good agreement with the experimentally observed color distributions in the fully relaxed bilayers (see **fig. S6** in Supplementary Materials for the simulated color patterns).

Comparing **Fig. 2** (**A and C**), increasing the CLCE film thickness $H_f$ from 160 μm to 206 μm (by dropping different amounts of CLCE ink onto the LCE substrate during fabrication) while keeping the same $\lambda_{pr}$ and $E_f/E_s$ results in an increased wrinkle wavelength at the fully relaxed state. However, the wrinkling mode remains the same "doubling" pattern, preserving the underlying strain distribution. Consequently, the resulting color distribution remains unchanged: the initially flat green surface transforms into blue at the wrinkle peaks and orange at the valleys. This effect is illustrated in **movie S2** (Supplementary Materials), which captures the stretching-relaxing process of CLCE-LCE bilayers with varying $H_f$. A further comparison is provided in **fig. S7**, showing wrinkle patterns for two different $H_f$. **Figure 2D** quantifies the dependence of wavelength and amplitude on the CLCE film thickness in fully relaxed CLCE-LCE bilayers. Both features increase linearly with increasing $H_f$, in strong agreement with FEA predictions.

The pre-stretch ratio $\lambda_{pr}$ of the LCE substrate plays a critical role in shaping the wrinkle morphology and modulating the wavelength. Comparing **Fig. 2** (**A** and **E**), increasing the substrate pre-stretch ratio to $\lambda_{pr} = 1.90$ results in a reduced wavelength and an increased amplitude at the fully relaxed state, producing sharper and more compact wrinkle patterns. This behavior is illustrated in **movie S3** (Supplementary Materials), which captures the stretching-relaxing process of CLCE-LCE bilayers with varying pre-stretch levels. **Figure S8** further compares multiple bilayers with different $\lambda_{pr}$. **Figure 2F** presents the dependence of wavelength and amplitude on the substrate pre-stretch ratio in fully relaxed CLCE-LCE bilayers. As $\lambda_{pr}$ increases, the wavelength decreases while the amplitude increases, consistent with FEA predictions. The color response of the bilayer is also tunable under partial relaxation. At a partially relaxed state ($\lambda = 1.45$), the surface exhibits a distinct alternating color pattern of orange and blue due to the exposed peak-valley structure. However, upon full relaxation ($\lambda = 1$), the wrinkle peaks begin to occlude the valleys from the top view, resulting in a dominant blue appearance across the surface. This demonstrates a strategy for dynamic color modulation by tuning the degree of relaxation in the bilayer system.

The bilayer modulus ratio $E_f/E_s$ is a key parameter governing the wrinkling mode, wrinkle wavelength, and color distribution, and is tuned by adjusting the chemical composition of the CLCE films (see **fig. S9** in Supplementary Materials for stress-strain curves of CLCE films with varying moduli). Comparing **Fig. 2** (**A** and **G**), increasing the $E_f/E_s$ to 95.73 leads to a distinct "ridging" pattern, accompanied by increased wavelength and amplitude at the fully relaxed state. The associated color response also changes: the initially flat green surface evolves into cyan at the wrinkle peaks and yellow at the valleys, which is attributed to the reduced strain variation across the CLCE film surface. This behavior is illustrated in **movie S4** (Supplementary Materials), which captures the stretching-relaxing process of CLCE-LCE bilayers with varying $E_f/E_s$. Further comparisons in **fig. S10** reveal a distinct "creasing" pattern at a lower $E_f/E_s = 11.69$. **Figure 2H** summarizes the quantitative relationship between $E_f/E_s$ and the wrinkle features in fully relaxed CLCE-LCE bilayers. Both the wavelength and amplitude increase with higher $E_f/E_s$, aligning well with FEA predictions.

These observations demonstrate that tuning CLCE film thickness, substrate pre-stretch ratio, and bilayer modulus ratio offers powerful control over both the onset and evolution of surface wrinkling, enabling access to a rich spectrum of morphological modes, including creasing, doubling, and ridging, with programmable wavelengths and amplitudes. This approach provides a highly versatile design framework for engineering CLCE-LCE bilayers with customizable and reconfigurable surface textures and colors.

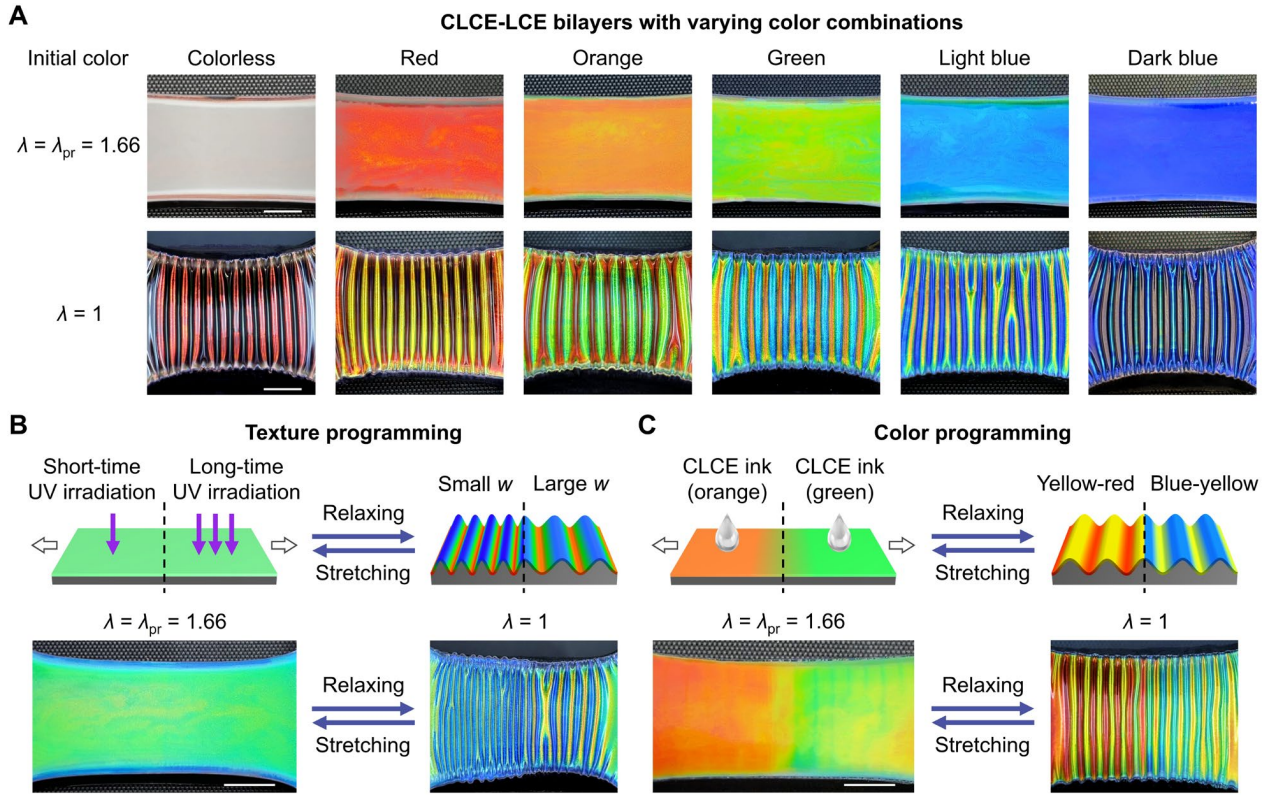

**Fig. 3. Tunable color combinations and programmable texture and color modulation.** (**A**) Images of CLCE-LCE bilayers with varying initial colors at the stretched state ($\lambda = 1.66$) and varying color combinations at the fully relaxed state ($\lambda = 1$). The bilayers have different initial colors – colorless, red, orange, green, light blue, and dark blue. (**B**) Schematics and images showing spatial texture programming achieved by locally modulating the modulus via controlled UV exposure. (**C**) Schematics and images showing spatial color programming achieved by tuning the initial composition of the CLCE ink. Scale bars: 5 mm.

## Tunable color combinations and programmable texture and color modulation

In the previous section, CLCE-LCE bilayers that transition from an initial green color when flat to an alternating orange-blue color when wrinkled are shown. Notably, the initial color of the CLCE film can be controlled across the entire visible spectrum by varying the amount of the chiral liquid crystal monomer added, which determines the initial pitch of the periodic helical nanostructures. As demonstrated in **Fig. 3A**, from left to right, colorless, red, orange, green, light blue, and dark blue stretched flat bilayers ($\lambda = \lambda_{pr} = 1.66$) are shown, and the lower row displays wrinkle patterns at the fully relaxed state ($\lambda = 1$). By controlling the initial color of the CLCE film, bilayers with different color combinations are realized. For example, a bilayer with an initially colorless CLCE film transforms into a wrinkle pattern with red peaks and black valleys, while an initially orange bilayer produces a wrinkle pattern with green peaks and red valleys (see **movie S5** in Supplementary Materials for the stretching-relaxing process of CLCE-LCE bilayers with varying color combinations). By cooperatively leveraging the programmable substrate pre-stretch and bilayer modulus ratio, the strain distribution within the CLCE film can be accurately modulated for combinations of all colors in the visible spectrum on the textured surface.

The CLCE-LCE bilayer's versatile programmability enables the integration of multiple surface textures and color modes within a single system. As demonstrated in **Fig. 3B**, spatial texture programming is achieved by locally controlling the modulus via controlled UV exposure. A green CLCE film is first deposited onto a pre-stretched LCE substrate. During UV curing, a mask selectively blocks specific regions of the bilayer, creating spatial variation in UV dosage. In the example shown, the left half of the CLCE film is exposed to UV light for 1 s, while the right half receives 10 s of exposure. These different curing times result in a stiffer CLCE film on the right, leading to a higher local modulus ratio. Upon full relaxation, the left half forms wrinkles with a smaller wavelength compared to the right, demonstrating that localized UV curing provides a powerful strategy for encoding spatially varying surface textures (see **movie S6** in Supplementary Materials for the stretching-relaxing process).

Similarly, spatial color programming on the textured surface is achievable by tuning the initial composition of the CLCE ink. As shown in **Fig. 3C**, orange- and green-colored CLCE inks are selectively deposited onto the left and right halves of a pre-stretched LCE substrate. After solvent evaporation and UV curing, the bilayer retains the distinct orange and green colors in their respective regions. Upon full relaxation of the bilayer, each half exhibits a unique wrinkle-induced color pattern: the left half has wrinkles colored yellow-red (peak-valley), while the right half has wrinkles colored blue-yellow (peak-valley) (see **movie S6** in Supplementary Materials). This approach demonstrates that spatial chemical patterning of the CLCE film enables programmable region-specific color responses on a textured surface, offering an additional degree of freedom for creating multifunctional, spatially encoded optical materials.

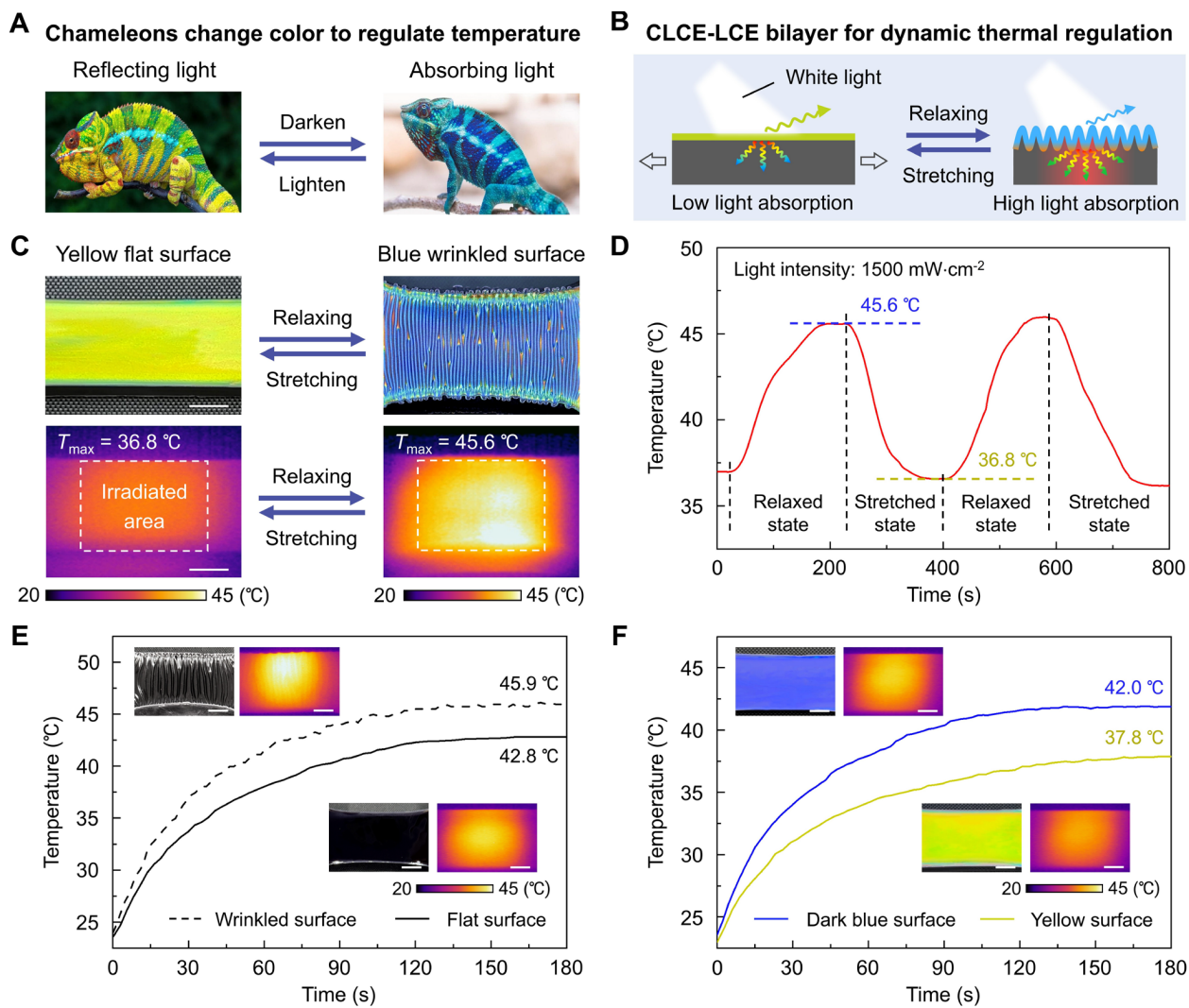

**Fig. 4. CLCE-LCE bilayer as smart skin for dynamic thermal regulation.** (**A**) A chameleon can regulate its body temperature by changing skin color. (**B**) Mechanism schematics of the CLCE-LCE bilayer for dynamic thermal regulation. (**C**) Images of a CLCE-LCE bilayer with an initial yellow color at the stretched state ($\lambda = 2$) and with dark blue wrinkles at the fully relaxed state ($\lambda = 1$), and corresponding IR images under white light irradiation. (**D**) Maximum surface temperature of the CLCE-LCE bilayer for dynamic thermal regulation versus irradiating time. Maximum surface temperature of (**E**) flat and wrinkled and (**F**) yellow and dark blue CLCE-LCE bilayers versus irradiating time. Scale bars: 5 mm.

## CLCE-LCE bilayer as smart skin for dynamic thermal regulation

Chameleons are well known for dynamically regulating their body temperature through skin color changes (**Fig. 4A**) (*39*). Darker skin tones increase light absorption and raise body temperature, while lighter colors reflect more light, reducing thermal gain. In addition, many biological systems also exploit surface texture changes to further enhance thermal regulation. For example, animals such as hedgehogs, birds, or certain desert lizards can raise or lower surface structures (like fur, feathers, or scales) to alter their exposed surface area, insulation, or light interaction, effectively modulating heat transfer through both geometry and optics. Inspired by these biological strategies, the CLCE-LCE bilayer system combines color change and texture modulation in a single material platform, enabling a smart skin for more efficient and responsive thermal regulation. As shown in **Fig. 4B**, a CLCE-LCE bilayer is capable of active thermal regulation by switching between a stretched, flat yellow state and a relaxed, wrinkled dark blue state, modulating both surface texture and color to control light absorption. **Figure 4C** demonstrates this concept using a CLCE-LCE bilayer with $H_f = 70$ μm, $\lambda_{pr} = 2$, and $E_f/E_s = 23.87$. When exposed to white light (1500 mW·cm$^{-2}$) for 3 min in the stretched flat state, the bilayer reaches a steady-state surface temperature of 36.8 °C, as captured by IR imaging. Upon relaxation, the surface transforms into a dark blue, wrinkled configuration, and the maximum temperature increases to 45.6 °C under the same illumination. This temperature variation demonstrates the synergistic thermal modulation enabled by combined color and topographical change. **Figure 4D** presents the reversible temperature switching behavior during cyclic relaxing and stretching of the CLCE-LCE bilayer, confirming the system's high repeatability and reliability in dynamic thermal tunability through coupled surface texture and color modulation (see **movie S7** in Supplementary Materials).

The decoupled individual contributions of surface texture and color are examined by two control bilayer systems, each with identical geometric parameters. **Figure 4E** shows an LCE bilayer with a colorless LCE film on top of a black LCE substrate, which is capable of forming wrinkles without any color change. Under 3 min of white light exposure, the wrinkled surface heats up to 45.9 °C, compared to 42.8 °C for the flat configuration. This increase is attributed to the wrinkled structure increasing the effective irradiated surface area, thereby enhancing light absorption (*40*). **Figure 4F** presents bilayers with uniform surface colors. The dark blue bilayer reaches a temperature of 42.0 °C, which is significantly higher than the yellow bilayer (37.8 °C), due to the stronger light absorption of darker color. Note that the thickness change of the bilayer during the stretching-relaxing has negligible effect on temperature variation (see **fig. S11** in Supplementary Materials). The combined modulation of texture and color enables a wider and more effective range of temperature control than using either mechanism alone. By actively tuning both surface wrinkles and color, the bilayer functions like a smart skin, adapting on demand to its environment by absorbing more heat in cold conditions and reflecting more heat in warm conditions. This reversible and responsive behavior closely mimics the biological thermal regulation.

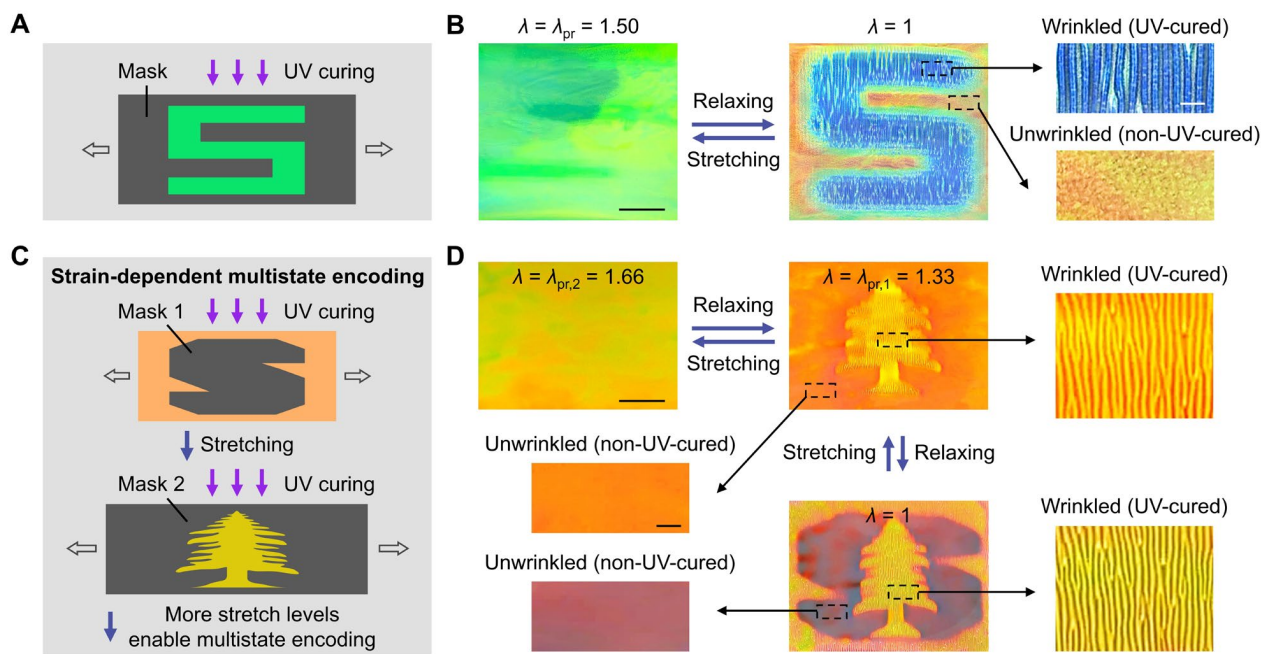

**Fig. 5. CLCE-LCE bilayer for multistate information encoding.** (**A**) Schematic and (**B**) experimental images demonstrating information encoding using mask-assisted selective UV curing. The exposed "S" pattern is cured at the stretched state ($\lambda = \lambda_{pr} = 1.50$). (**C**) Schematics and (**D**) experimental images demonstrating multistate information encoding. The area outside the shaded "S" pattern is first cured at the stretched state ($\lambda = \lambda_{pr,1} = 1.33$), followed by curing of the exposed "tree" pattern at the stretched state ($\lambda = \lambda_{pr,2} = 1.66$). Scale bars: 10 mm for regular views and 1 mm for enlarged views.

## CLCE-LCE bilayer for multistate information encoding

Programmable encoding of information offers a powerful strategy for designing intelligent materials capable of dynamic display and encryption. Here, we design the CLCE-LCE bilayer as an erasable information display platform. By selectively controlling UV-cured and non-UV-cured regions within the bilayer, we program spatially defined wrinkle patterns that serve as tunable and reversible carriers of visual information. As shown in **Fig. 5A**, a green CLCE film is formed on a pre-stretched LCE substrate (40 mm × 40 mm × 4 mm, $\lambda_{pr} = 1.50$) and selectively cured through an "S"-shaped photomask. The exposed "S" region forms a stiff, fully-crosslinked CLCE layer, while the shaded region remains soft and lightly crosslinked. As illustrated in **Fig. 5B**, the stretched bilayer initially shows a uniform green color. During the strain relaxation process, the cured "S" region develops blue wrinkles, while the surrounding uncured area remains smooth and shifts to orange, revealing a visible "S" pattern. This process is fully reversible, and the "S" pattern disappears when the bilayer is stretched again (see **movie S8** in Supplementary Materials).

This concept is further extended to strain-dependent multistate information encoding via multistep selective UV curing at different stretch levels, which enables hierarchical control over the wrinkle patterns. Using a two-state information-encoding CLCE-LCE system as an example, an orange CLCE film is first selectively cured under an "S"-shaped photomask at $\lambda = \lambda_{pr,1} = 1.33$ (**Fig. 5C**). After further stretching to $\lambda = \lambda_{pr,2} = 1.66$, the bilayer turns yellow, and a second "tree" pattern is cured using a new mask. As illustrated in **Fig. 5D**, the bilayer initially appears yellow without visible patterns. Upon relaxation from $\lambda = 1.66$, the wrinkled "tree" pattern appears first at the intermediate state ($\lambda = 1.33$), while the surrounding area remains smooth and turns orange. Further relaxation to $\lambda = 1$ results in wrinkling outside the "S" region, which highlights the smooth red "S" pattern through morphological and color contrast (see **movie S8** in Supplementary Materials). This multistep selective UV curing strategy enables spatially and temporally programmable information encoding, which establishes a versatile platform for information storage, encryption, and dynamic optical display.

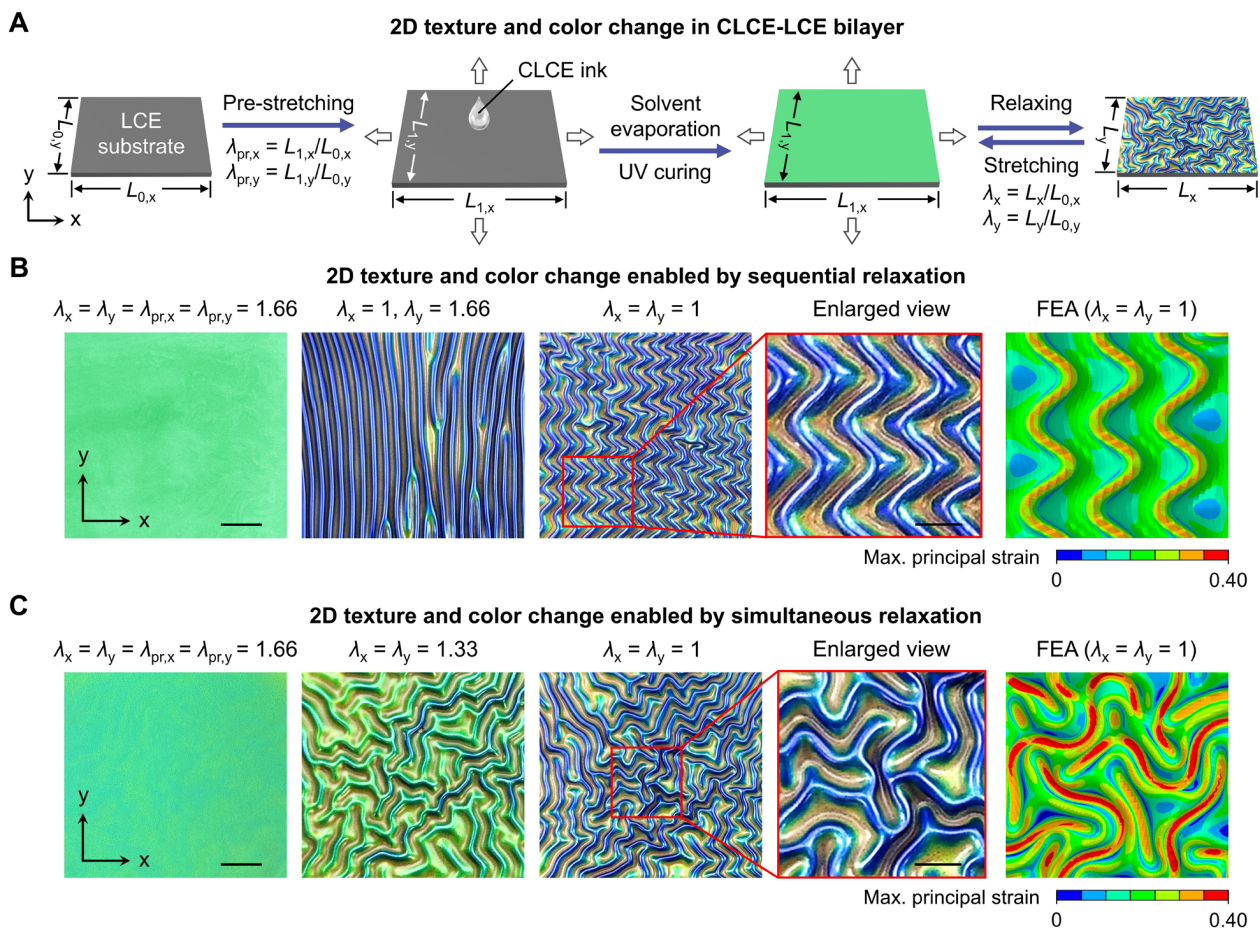

**Fig. 6. 2D texture and color modulation of CLCE-LCE bilayer.** (**A**) Schematics of the fabrication process of CLCE-LCE bilayer showing 2D texture and color changes during wrinkling. Experimental images showing the evolution of 2D texture and color in CLCE-LCE bilayers during the (**B**) sequential and (**C**) simultaneous strain relaxation processes, and corresponding FEA images at the fully relaxed state ($\lambda_x = \lambda_y = 1$). The bilayers have $H_f = 160$ μm, $\lambda_{pr,x} = \lambda_{pr,y} = 1.66$, and $E_f/E_s = 32.46$. Scale bars: 5 mm for regular views and 2 mm for enlarged views.

## 2D texture and color modulation

Controlling 2D surface texture and color modulation expands the design space for multifunctional optical materials. When biaxial pre-stretching is applied to a CLCE-LCE bilayer, diverse complex 2D wrinkle patterns can be generated by adjusting the relative strain levels between the two planar directions (*41*). **Figure 6A** illustrates the fabrication process of a biaxially pre-stretched CLCE-LCE bilayer that exhibits 2D texture and color changes upon strain relaxation. An LCE substrate with original lengths of $L_{0,x}$ and $L_{0,y}$ along x- and y-directions is biaxially pre-stretched to $L_{1,x}$ and $L_{1,y}$, respectively. The pre-stretch ratios along x- and y-directions are defined as $\lambda_{pr,x}$ and $\lambda_{pr,y}$. CLCE ink is dropped onto the pre-stretched LCE substrate. After solvent evaporation and UV curing, simultaneous 2D texture and color changes are observed upon full relaxation. **Figure 6 (B and C)** present the wrinkling processes of bilayers upon sequential and simultaneous relaxation. Both bilayers are relaxed from a stretched state ($\lambda_x = \lambda_y = \lambda_{pr,x} = \lambda_{pr,y} = 1.66$). Under sequential relaxation, initial 1D wrinkles form along the first relaxed axis and then evolve into an ordered zigzag pattern, with the wrinkle peaks and valleys exhibiting blue and red shifts, respectively (**Fig. 6B**). In contrast, simultaneous relaxation results in an irregular maze-like wrinkle pattern with increasing wrinkle density and similar color shifts throughout the process (**Fig. 6C**). In both cases, the FEA results match well with the experimental results. **Figure S12** shows two additional bilayers that underwent simultaneous relaxation with the same $\lambda_{pr,x}$ but different $\lambda_{pr,y}$. Compared to the bilayer in **Fig. 6C**, these bilayers exhibit less disordered wrinkle patterns with a y-direction preference and smaller color shifts due to reduced mismatch strains along y-direction. These findings demonstrate that 2D strain programming enables precise and independent control of surface morphology and color along each direction, unlocking greater design freedom for generating complex and programmable optical patterns.

## DISCUSSION

In summary, we have designed and fabricated a bioinspired CLCE-LCE bilayer with simultaneous and reversible modulation of surface texture and structural color via controllable wrinkling. Tunable wrinkle morphologies and color combinations are achieved through precise control over four parameters: CLCE film thickness, LCE substrate pre-stretch ratio, bilayer modulus ratio, and the initial structural color of CLCE film. FEA simulations have been conducted to model the wrinkle patterns and color distributions of the bilayer, which agree very well with the experimental results. Spatial programming of surface texture and color is also realized to produce diverse patterns by localized UV curing and spatial chemical patterning of the CLCE film. Potential applications in dynamic thermal regulation and multistate information encoding are demonstrated, highlighting the versatility of the CLCE-LCE bilayer system. 2D texture and color modulation is further explored via biaxial pre-stretching. It is anticipated that this work will establish a versatile platform for developing advanced multifunctional materials with customizable texture and color as well as deepen the understanding of the interplay between mechanical instabilities and structural color modulation in soft material systems.

## MATERIALS AND METHODS

### Fabrication of LCE substrate

To fabricate the LCE substrate, the di-acrylate liquid crystal monomer 1,4-Bis-[4-(3-acryloyloxypropyloxy)benzoyloxy]-2-methylbenzene (RM257, Kindchem (Nanjing) Co., Ltd, China) was dissolved in toluene (50.0 wt%) at 80 °C for 5 min. After the mixture was cooled to room temperature, the di-thiol chain extender 2,2-(ethylenedioxy) diethanethiol (EDDET, Sigma Aldrich, USA) (22.0 wt%), the tetra-thiol crosslinker pentaerythritol tetrakis(3-mercaptopropionate) (PETMP, Sigma Aldrich, USA) (6.0 wt%), and the black dye (nigrosine, Sigma Aldrich, USA) (0.1 wt%) were introduced to the mixture. The black dye was used to enhance the visibility of the structural color. The mixture was then stirred for 3 min with a magnetic stir bar for homogenization, and the catalyst dipropylamine (DPA, Sigma Aldrich, USA) (0.8 wt%) was added to the mixture. After stirring and degassing, the mixture was poured into a polydimethylsiloxane (PDMS) mold to undergo Michael addition reaction at room temperature for 12 h and was heated at 80 °C for 1 h to evaporate the toluene. After demolding, the lightly crosslinked stretchable LCE substrate was obtained. All weight ratios are given relative to RM257.

### Fabrication of CLCE ink

The fabrication of CLCE ink formulated to produce green color is described below: RM257 and the di-acrylate chiral liquid crystal monomer (3$R$,3a$S$,6a$S$)-hexahydrofuro[3,2-b] furan-3,6-diyl bis(4-(4-((4-(acryloyloxy)butoxy) carbonyloxy) benzoyloxy)benzoate) (LC756, Kindchem (Nanjing) Co., Ltd, China) (5.2 wt%) were dissolved in toluene (80.0 wt%) at 80 °C for 5 min. After the mixture cooled to room temperature, EDDET (27.0 wt%) and PETMP (4.0 wt%) were introduced to the mixture and stirred for 3 min for homogenization. Then, the photoinitiator Irgacure 819 (Sigma Aldrich, USA) (2.0 wt%) and DPA (0.4 wt%) were added to the mixture. After stirring for 1 min, the CLCE ink formulated to produce green color was obtained. All weight ratios are given relative to RM257. The fabrication of CLCE inks for different demonstrations followed the same procedures but with different component contents (see **table S1** in Supplementary Materials).

**Fabrication of CLCE-LCE bilayer**

To fabricate the CLCE-LCE bilayer, CLCE ink was dropped onto a uniaxially (for 1D texture) or biaxially (for 2D texture) pre-stretched LCE substrate, which was exposed to air for 2 h at room temperature. The CLCE-LCE bilayer was obtained after being irradiated by UV light (385 nm) for 10 s to cure the CLCE film and bond it with the LCE substrate. Upon strain relaxation, both wrinkling and color change were observed.

## Acknowledgments

**Funding:** The authors acknowledge the support from the Army Research Office (ARO) ECP Award W911NF-23-1-0176 and National Science Foundation (NSF) Career Award CMMI-2145601 and CMMI-2318188.

**Author contributions:** R.R.Z. conceived the research and supervised the study. X.Y. performed the experiments and data analysis. J.S. and W.H. contributed to the FEA simulations. X.Y., J.S., and R.R.Z. wrote the original draft. All authors edited the original draft.

**Competing interests:** The authors declare that they have no competing interests.

**Data and materials availability:** All data needed to evaluate the conclusions in the paper are present in the paper and/or the Supplementary Materials.